\documentclass{article}

\usepackage{amsmath, amsthm, amsfonts}
\usepackage{graphicx}
\usepackage{xcolor}

\newtheorem{theorem}{Theorem}[section]
\newtheorem{corollary}{Corollary}

\newtheorem{remark}{Remark}

\newcommand{\be}[1]{\begin{equation}\label{#1}}
\newcommand{\ee}{\end{equation}}

\title{Dynamical system describing cloud of particles in relativistic and non-relativistic framework}
\author{R. Sta\'nczy \\ Uniwersytet Wrocławski\\ D. Bors \\ University of Lodz}

\begin{document}
\maketitle

\begin{abstract}%
We consider fairly general class of dynamical systems under the assumptions guaranteeing the existence of Lyapunov function around some nontrivial stationary point. Moreover, the existence of heteroclinic trajectory is proved motivated by integrated densities approach to some astrophysical models of self-gravitating particles both in relativistic and non--relativistic frameworks. Finally, with the aid of geometric and topological reasoning we find the upper bounds for this trajectory yielding the critical mass--radius theorem for the astrophysical model.
\end{abstract}

\maketitle

\section{Intoduction}
Consider, for given smooth functions $a,b$ defined on some interval, the system
\begin{eqnarray*}
x'=y-x\,,\\
y'=a(x)y-b(x)y^2\,.
\end{eqnarray*}
Under some appropriate structural assumptions on $A'=a,B'=b$ with $z>0$ being the zero of the second equation $a(z)=zb(z)$ we shall prove that
$$
V=zB(x)-A(x)+y-z-z\log (y/z)\,.
$$
is the Lyapunov function of the system, providing with some extra assumptions the existence of heteroclinic trajectory joining $(0,0)$ with $(z,z)$. Moreover, by geometric analysis combined with topological reasoning, we estimate $x$ values of the heteroclinic trajectory. The dynamical system considered above arises from some astrophysical model. Thus the critical mass-radius theorem follows, since the dynamical variables correspond to the integrated density of mass variables. The key role play the following examples: $a(x)=2-x$ and $b(x)=0$ for nonrelativistic and $a(x)=\frac{2-3x}{1-x}, \,b(x)=\frac{1}{1-x}, x<1$ for relativistic case. 
 
\section{Astrophysical motivation}
In recent years due to observational achievements of stable trajectories in the centre of Milky Way, see \cite{Sag} observed by ESO including S2 and unstable ones to account for Hills mechanism made by Koposov et al. for by S5 for HVS1, cf. \cite{S5K}  there have been suggested some dark matter models to explain it, see  P.H. Chavanis \cite{C}, as an alternative to black hole models provided by Christodoulou  \cite{D}. One of the questions that arises is the critical mass--radius relation that these dark matter models allow for. In this paper we analyze dynamical system for the integrated density of mass encompassing both nonrelativistic model described by stationary radially symmetric solutions to Smoluchowski--Poisson equation and relativistic one by Tolman--Oppenheimer--Volkoff equation as a static and symmetric form of Einstein equation.

\section{Smoluchowski--Poisson equation for nonrelativistic particles}
 Smoluchowski--Poisson eq. for a cloud of self-gravitating particles
  $$
  \varrho_t=\nabla \cdot \left(P'(\varrho\theta^{-d/2})( \theta P'(\varrho\theta^{-d/2})\nabla \varrho + \varrho\nabla(\Delta)^{-1}\varrho)\right)
  $$
with equation of state binding the density $\varrho$ and the temperature $\theta$ with the pressure $p$ via relation
  $$
  p=\theta^{d/2+1}P(\varrho\theta^{-d/2})
  $$
  for $P={\rm Id}$ yields $p=\theta \varrho$. If we look for radial, stationary solutions in integrated density variables, cf. \cite{BHN} or \cite{BS, BSJ} we end up with the system of ODE's $x'=y-x, y'=(2-x)y$ and look for the heteroclinic trajectory linking $(0,0)$ with $(2,2)$ to find the density profile for given mass and radius of the system.
    
\section{Tolman--Oppenheimer--Volkoff for realtivistic model of particles}
Tolman--Oppenheimer--Volkoff equation follows from for static, symmetric form of Einstein equations and reads 
$$\frac{dp}{dr}=- \frac{G(\varrho(r)c^2+p(r))(m(r)c^2+4\pi r^3 p(r)}{rc^2(rc^2-2Gm(r))}\,.$$
Then the following change of variables $$rc^2x(\log r)=2Gm(r), c^2y(\log r)=8G\pi r^2\varrho(r)$$ with mass variable equal to $m(r)=4\pi \int_0^r s^2\varrho(s)ds$
  leads to the system $$x'=y-x,\;y'=2y-y(x+y)/(1-x)$$ if $p=c^2\varrho$ with Lyapunov function $$V=2y-\log(y)-6x-3\log (1-x)+C\,.$$ For details see \cite{BSM, BSA} and also see the next section for more general approach.

\section{Lyapunov function}

Consider, for given smooth functions $a,b$ defined on some interval, the system
\begin{eqnarray*}
x'=y-x\,,
y'=a(x)y-b(x)y^2\,.
\end{eqnarray*}
\begin{theorem}
Assume that there exist $z>0$ and a nonnegative, continuous function $r$ such that
\be
zzb(x)-a(x)=-r(x)(z-x)\,.
\ee
Denote by $A,B$ primitives of $a,b$ resp., i.e. $A'=a$ and $B'=b$ requiring additionally that $A(z)=0$ and $B(z)=0$. Then the Lyapunov function reads
$$
V=zB(x)-A(x)+y-z-z\log (y/z)\,.
$$ 
\end{theorem}
{\bf Proof.}
Then the first equation $x'=y-x$ multiply by $zb(x)-a(x)$ while equation $y'=a(x)y-b(x)y^2$ by $1-z/y$. Thus by $zb(x)-a(x)=-r(x)(z-x)$ we get $$\left( zB(x)-A(x)+y-z\log (y)\right)'=-b(x)(y-z)^2+(zb(x)-a(x))(z-x)\le 0\,.$$
Indeed, it is nothing else than $(y-x)(zb(x)-a(x))+(a(x)y-b(x)y^2)(1-z/y)\,.$ $\qed$

\section{Examples}
Examples coming from astrophysical models of self-attracting particles are the following, with some constant $C$ which might differ from line to line:
\begin{itemize}
  \item $a(x)=2-x, b(x)=0, z=2, A(x)=2x-x^2/2-2$ and $$2V=(x-2)^2+2y-4\log y+C$$ nonrelativistic $a-zb=2-x$ 
  \item $a(x)=\frac{2-3x}{1-x}, b(x)=\frac{1}{1-x}, A(x)=3x+\log(1-x)+C, B=-\log 2-\log (1-x)$
  $$2V=-6x-3 \log(1-x)+2y- \log y+2-4\log 2$$ relativistic $z=1/2$ where
  $2(1-x)(zb-a)=1-4+6x=6(x-1/2)$
  \item $a(x)=\frac{2-24\pi x}{1-8\pi x}, b(x)=\frac{8\pi x}{1-8 \pi x}, A(x)=3x+\frac{1}{8\pi} \log(1-8\pi x)-\frac{3}{16\pi}+\frac{1}{8\pi}\log 2, B(x)=-\log (1-8\pi x)-\log 2$
  \begin{figure}[h]
\includegraphics[height=5.7cm]{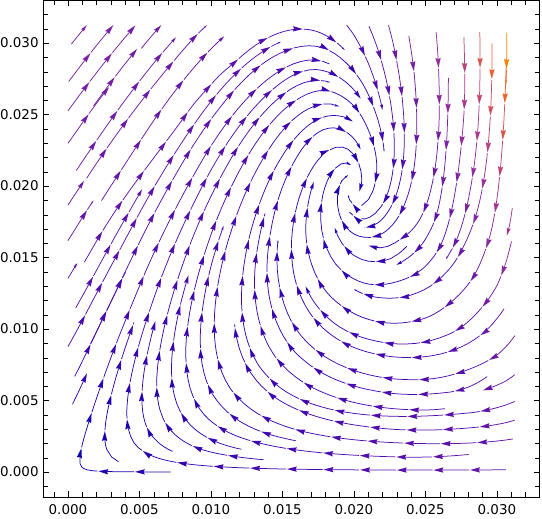}
\includegraphics[height=5.7cm]{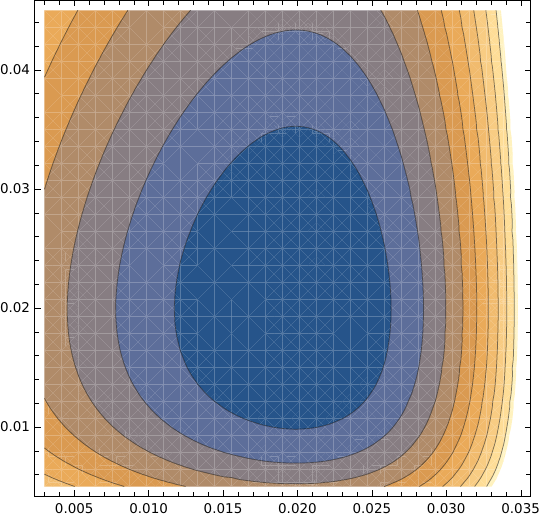}
\caption{Phase portrait on the left. On the right Lyapunov function level sets. }
\end{figure}
  $$16\pi V=-48\pi x-3 \log(1-8\pi x)+16\pi y- \log y-\log (128\pi ) +2$$ 
  relativistic $z=1/{16 \pi}$ and $w=\frac{1}{24\pi}$ where
  $2(1-8\pi x)(zb-a)=3(16\pi x-1)$

  % \begin{figure}
%\includegraphics[height=7cm]{hfx.pdf}
%\end{figure}
  \item $a(x)=2-\frac{1+\kappa}{2\kappa}\frac{x}{1-x}=\frac{4\kappa-(1+5\kappa)x}{2\kappa(1-x)}$, $b(x)=\frac{1+\kappa}{2}\frac{1}{1-x}$ hence
  $$2\kappa(1-x)(a(x)-xb(x))=4\kappa-x(\kappa^2+6\kappa+1)$$ 
  thus
  $z=\frac{4\kappa}{(\kappa+1)^2+4\kappa}\,.$
  $$
V=2y-(5+1/\kappa)x-2x_k \log \left(y(1-x)^{\delta_\kappa}\right)+C_\kappa
$$ 
where the exponent $\delta_\kappa$ is defined by
$$
8\kappa^2\delta_\kappa=(5\kappa+1)(\kappa+1)^2
$$
while the constant $C_\kappa=(3+1/\kappa)x_\kappa+2x_\kappa \log \left( x_\kappa(1-x_\kappa)^{\delta_\kappa}\right)\,.$

\end{itemize}
\section{Upper bound for heteroclinic trajectory}

\begin{theorem}
Assume $b(x)\ge 0$, $a(x)$ are $C^1$ and that there exist $r(x)\ge 0$ and constants $(a(0)+1)w>z\ge w>0$ such that
\begin{itemize}
\item $zb(x)-a(x)=-r(x)(z-x)$\,,
\item $(a(0)+1)wb(w)=a(w), a(0)>0$\,,
\item $(a(0)+1)wb(x)\ge a(x)-a(0), x\le w$\,.
\item $a'(x)-b'(x)y<0, \,y\in [z,(a(0)+1)w],\, x\in [w,z]$
\end{itemize}
Then the heteroclinic trajectory joining $(0,0)$ with $(z,z)$ can be estimated in the $x$ variable by
$$
X=H^{-1}\left( (a(0)+1)w-z-z\log ((a(0)+1)w/z)\right)\,,
$$
where $H(x)=zB(x)-A(x), x\ge z$.
\end{theorem}
\noindent {\bf Proof.} Analysis of stability of stationary points has been postponed to separate section. Just recall that $(0,0)$ is saddle and $(z,z)$ is asymptotically stable. Let us note that when $y=(a(0)+1)x$ then from the third assumption it follows that for $x\le w$ we have
$$\frac{y'}{x'}=\frac{(a(x)y-b(x)y^2}{y-x}=(a(x)-b(x)(a(0)+1)w)(a(0)+1)/a(0)\le a(0)+1\,.$$
Therefore we know that as long as $x\le w$ the vector field at the line $y=(a(0)+1)x$ is directed below this line. Moreover, if $0\le y=x\le z$ then $x'=0$ and $y'=a(x)y-b(x)y^2=y(r(z-x)+b(z-y))\le 0$ so the vector field is directed upwards on this segment. Finally, from the last assumption it follows, that there exists unique curve $(x(y),y)$ defined by $a(x)-b(x)y=0$ joining $(w,(a(0)+1)w$ with $(z,z)$ with nonincreasing function $x=x(y)$ defined on the interval $[z,(a(0)+1)w]$ mapping it onto $[w,z]$.

Looking for the intersection of the unstable direction $y=(a(0)+1)x$ and the isocline $yb(x)=a(x)$ by the second assumption we get a solution at $x=w$ satisfying
$$
(a(0)+1)wb(w)=a(w)\,.
$$
Finally we establish the unique maximal solution $x=X$ of the equation
$$
V(x,z)=V(z,(a(0)+1)w)
$$
since at $z$ the level set of the $V(x,z)=V(z,(a(0)+1)w)$ yields maximal value of $x=X$. To be more specific the above equation leads due to form of the Lyapunov function
$$
zB(x)-A(x)=(a(0)+1)w-z-z\log ((a(0)+1)w/z)
$$
Since the function $H(x)=zB(x)-A(x), x\ge z$ is monotone in its domain due to the first assumption thus the heteroclinic trajectory enjoys the upper bound for $x$ values as $X=\max x$ defined by
$$
X=H^{-1}\left( (a(0)+1)w-z-z\log ((a(0)+1)w/z))\right)\,.
$$
\begin{corollary}
In the case $a(x)=\frac{2-3x}{1-x}$ and $b(x)=\frac{1}{1-x}$ we have the upper bound for the $x$ variable heteroclinic trajectory joining $(0,0)$ and $(1/2,1/2)$ as 
$$
2H(X)= \log z -2z+2(a(0)+1)w-\log ((a(0)+1)w)\,.
$$
where $2H(x)=-6x-3\log(1-x)+6z+3\log (1-z), x\ge 1/2$. Thus
$$
2X=2+W(-2\exp(-4/3-2/3 \, \log 2))=2+W\left( -2^{1/3}e^{-4/3} \right)\,.
$$
where $W$ is Lambert function, productlog, i.e. inverse of $(\cdot)\exp(\cdot)$ and $X<0.7$.
\end{corollary}

\begin{remark}
The pictures presented below come from \cite{BSM, BSA} with rescaled by factor $8\Pi$ version of the above corollary and are included here only for illustration, since they were generated for the relativistic example with $a(x)=\frac{2-24\pi x}{1-8\pi x}$ and $b(x)=\frac{8\pi x}{1-8 \pi x}$ note that $1/{8\pi}\sim 0.04$ while $1/{16\pi}\sim 0.02$ hence the nontrivial stationary point is around $(0.02,0.02)$ and the tangency of level set of Lyapunov function or intersection of tangent to unstable manifold and one of the isoclines both happen at around $y\sim 0.04$.
\end{remark}
\begin{figure}[h]
\includegraphics[height=5cm]{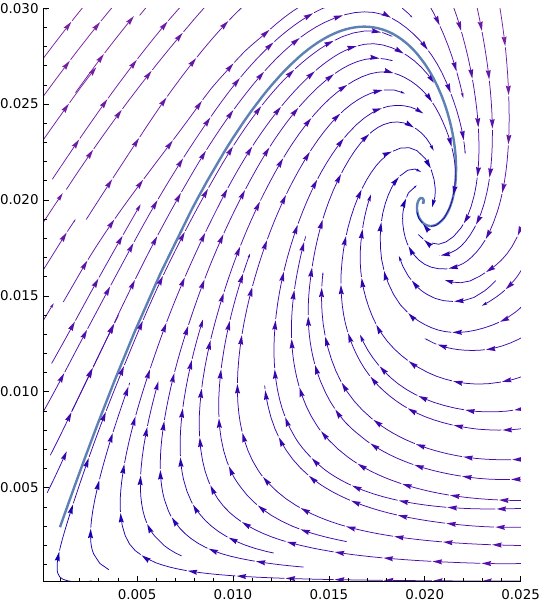}
\includegraphics[height=5cm]{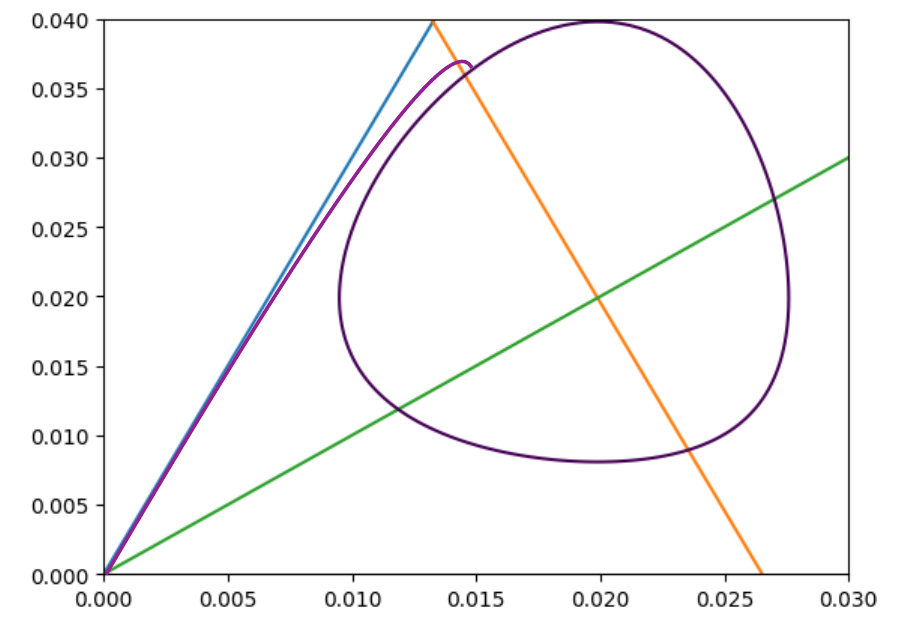}
\caption{Heteroclinic trajectory with phase portrait. On the right pic.: \textcolor{blue}{unstable tangent}, \textcolor{orange}{isocline y'=0}, \textcolor{olive}{isocline x'=0}, \textcolor{violet}{crucial Lyapunov level set}.}
\end{figure}

\begin{corollary}
In the case $a(x)=2-x$ and $b(x)=0$ we obtain the upper bound $X$ for the $x$ variable of the heteroclinic trajectory joining $(0,0)$ and $(2,2)$ as 
$$
X=H^{-1}\left( 4-2\log 3)\right)\,.
$$
where $2H(x)=(x-2)^2, x\ge 2$, whence $H^{-1}(y)=2+\sqrt{2y}$. Thus
$$
X=2+2\sqrt{2-\log 3}\,.
$$
\end{corollary}

\begin{corollary}
In the case $a(x)=2-\frac{1+\kappa}{2\kappa}\frac{x}{1-x}=\frac{4\kappa-(1+5\kappa)x}{2\kappa(1-x)}$, $b(x)=\frac{1+\kappa}{2}\frac{1}{1-x}$ we have
  $$2\kappa(1-x)(a(x)-xb(x))=4\kappa-x(\kappa^2+6\kappa+1)$$ 
  thus
  $$z=\frac{4\kappa}{(\kappa+1)^2+4\kappa}\,.$$. Moreover, we have 
$$zB(x)-A(x)=-\frac{1+5\kappa}{2\kappa}x-\frac{1+\kappa}{2\kappa}\frac{4\kappa+(1+\kappa)^3}{4\kappa+(1+\kappa)^2}\log (1-x)+D_\kappa$$
where $D_\kappa$ is chosen so that $A(z)=zB(z)$, i.e.
$$
D_\kappa=\frac{1+5\kappa}{2\kappa}z+\frac{1+\kappa}{2\kappa}\frac{4\kappa+(1+\kappa)^3}{4\kappa+(1+\kappa)^2}\log (1-z)\,.
$$
whence
$$
D_\kappa=\frac{2(1+5\kappa)}{4\kappa+(1+\kappa)^2}+\frac{1+\kappa}{2\kappa}\frac{4\kappa+(1+\kappa)^3}{4\kappa+(1+\kappa)^2}\log \left( \frac{(1+\kappa)^2}{4\kappa+(1+\kappa)^2} \right)\,.
$$
Moreover, solving
  $$
2\kappa(1-x)(a(x)-x(a(0)+1)b(x))=  4\kappa-x(3\kappa^2+8\kappa+1)=0
  $$
  gives $$w=4\kappa/(3\kappa^2+8\kappa+1)\,.$$ 
Hence, the upper bound for the $x$ variable of the heteroclinic trajectory joining $(0,0)$ and $(z,z)$ is equal to
$$
X=1+W\left( -\alpha_\kappa\exp\left( -\alpha_\kappa - (E_\kappa-D_\kappa) \frac{2\kappa}{1+\kappa}\frac{4\kappa+(1+\kappa)^2}{4\kappa+(1+\kappa)^3} \right)\right)/\alpha_\kappa\,.
$$
with
$$
\alpha_\kappa=\frac{1+5\kappa}{1+\kappa}\frac{4\kappa+(1+\kappa)^2}{4\kappa+(1+\kappa)^3}
$$
where $E_\kappa=(a(0)+1)w-z-z\log ((a(0)+1)w/z)$ and $W$ is Lambert function or productlog, i.e. inverse of $(\cdot)\exp(\cdot)$. 
\begin{figure}[h]
\includegraphics[height=3.6cm]{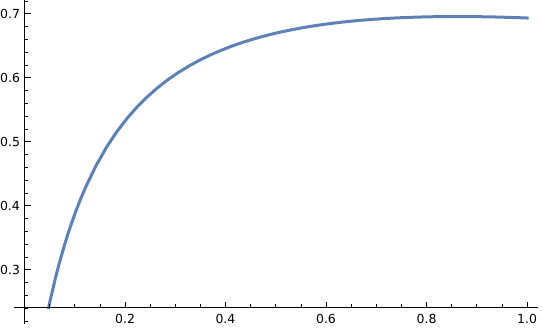}
\includegraphics[height=3.6cm]{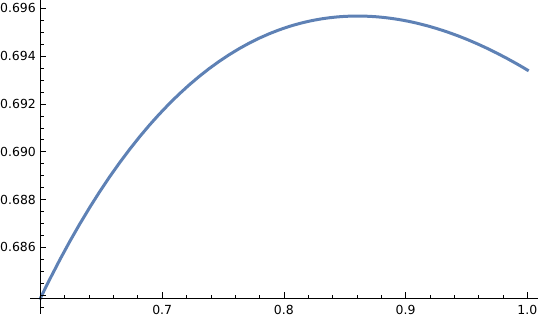}
\caption{Dependence of the bound $X$ of $x$ of the heteroclinic trajectory  on $\kappa$.}
\end{figure}

In fact
$$
E_\kappa=\frac{12\kappa}{3\kappa^2+8\kappa+1}-\frac{4\kappa}{4\kappa+(1+\kappa)^2}-\frac{4\kappa}{4\kappa+(1+\kappa)^2}\log \left( \frac{3\kappa^2+18\kappa+3}{3\kappa^2+8\kappa+1} \right)\,.
$$
Note that if $\kappa=1$ then $\alpha_\kappa=2$ and $E_\kappa=1/2-1/2\log 2$, $D_\kappa=3/2-3/2 \log 2$ with $\frac{2\kappa}{1+\kappa}\frac{4\kappa+(1+\kappa)^2}{4\kappa+(1+\kappa)^3}=2/3$ in agreement with the previous case, presented in corollary. One more value is important, i.e. $\kappa=1/3$ then $$X=1+\frac{25}{42}W\left(-8\cdot 3^{41/50} \cdot 7^{9/50} \cdot e^{-6/5}/25\right)$$ yielding $X<0.622$.
\end{corollary}

\begin{remark}
In fact in all the corollaries we have the existence of $x=x(y)$ as a smooth nonincreasing function of $y$ such that
$$
a(x(y))=yb(x(y))
$$ 
and satisfying the following conditions $x(z)=z, x((a(0)+1)w)=w, x(0)=x_0, a(x_0)=0$ for some $w\ge z\ge x_0 >0$. Thus $x:[0,(a(0)+1)w]\rightarrow [w,x_0]\ni z$ maps one interval into another reversing the order. We allow degenerate case $w=z=x_0$ like in our nonrelativistic example.
\end{remark}

\section{Analysis of stability of stationary solutions}
The linearization at $(0,0)$ yields the system $x'=y-x, y'=a(0)y$ and eigenvalues with the corresponding eigenvectors as $-1$ with $[1,0]$ and $a(0)$ with $[1,a(0)+1]$. Alternatively, to establish angle of the unstable manifold at $(0,0)$ and to establish heteroclinic connection between $(0,0)$ and $(z,z)$ we calculate
$$
I=\lim_{t\rightarrow-\infty}\frac{y}{x}=\lim_{t\rightarrow-\infty}\frac{y'}{x'}=\lim_{t\rightarrow-\infty}\frac{a(x)-b(x)y}{1-x/y}=a(0)/(1-1/I)
$$
hence
$$I=a(0)+1\,.$$
The linearisation at $(z,z)$ yields the following linear system $x'=y-x$ and $y'=(a'(z)z-b'(z)z^2)x+(a(z)-2b(z)z)y$. Hence if we recall relation $zb(x)=a(x)-r(x)(z-x)$ whence $r(z)=zb'(z)-a'(z)$ we get eigenvalues $2\lambda_{+/-}=-(1+a)-\sqrt{(1-a)^2-4zr}$ where square root works also for complex numbers. Hence if $4zr(z)=4z(zb'(z)-a'(z))>(1-a)^2$ then ${\mathcal Re}(\lambda_{+/-})=-(1+a(z))$ and if $4zr(z)=4z(zb'(z)-a'(z))\le (1-a)^2$ then ${\mathcal Re}(\lambda_{+/-})<-2a$. Therefore, both in the complex and real cases we have asymptotically stable solution provided $a(z)>-1$. 
\section{Examples}
As far as the stability for our examples are concerned:
\begin{itemize}
  \item $a(x)=2-x, b(x)=0, z=2$ and $4zr(z)=4z(zb'(z)-a'(z))> (1-a)^2$ is satisfied since $8>1$,
  \item $a(x)=\frac{2-3x}{1-x}, b(x)=\frac{1}{1-x}, (1-x)^2a'(x)=-1, (1-x)^2b'(x)=1$ whence $a(1/2)=1, b(1/2)=2, r(1/2)=6, a'(1/2)=-4, b'(1/2)=4$ and the condition to be verified is $12>0$ so the complex eigenvalues read $-1+i\sqrt{3}$ and $-1-i\sqrt{-3}$
  \end{itemize}

%\begin{figure}
%\includegraphics[height=5cm]{sol.pdf}
%\includegraphics[height=5cm]{pyttt.pdf}
%\end{figure}
  % \begin{figure}
%\includegraphics[height=7cm]{hfx.pdf}
%\end{figure}
  
%  \begin{figure}
%\includegraphics[height=11cm]{gkf.pdf}
%\end{figure}
%\end{itemize}

\section{Mass-radius ratio limits in astrophysical models}
Recall that $1$ stands for the definition of the rescaled Schwarzschild mass--radius ratio i.e. $$2GM/Rc^2<1$$ in the description of the black hole model with geometry defined by Schwarzschild metric. In our approach we follow the continuous model of accumulated mass $m(r)$ obeying TOV equation
$$
\frac{dp}{dr}=- \frac{G(\varrho(r)c^2+p(r))(m(r)c^2+4\pi r^3 p(r)}{rc^2(rc^2-2Gm(r))}\,.
$$
Apparently, also in our case the ratio should be bounded $2Gm(r)/rc^2<1$ and if we denote $M=m(R)$ then the same relation as in black hole model should hold for $M, R$. In this section we shall recall and provide new bounds for the rescaled mass radius ratio, i.e. $M/R$ rescaled by the factor $2G/c^2$. To be more specific we have the following estimates for $2GM/Rc^2$ as conclusions from the corollaries presented in the previous sections compared with known estimates:
  \begin{itemize}
  \item $8/9=0.97$ TOV, Buchdahl, Schwarzschild
  \item $12\sqrt{2}-16=0.95$ Bondi for $\varrho\ge 0$, \cite{B}
  \item $0.7<3/4=0.75$ new and now better than e.g. in Bors, Sta\'nczy for $c^2\varrho=p$, cf. \cite{BSM, BSA}
  \item $0.622$ our new estimate in the border case $c^2\varrho= 3p$ is better than $0.64$ obtained by Bondi provided that $c^2\varrho\ge 3p$
  \item $0.55$ this is some optimal numeric evaluation of heteroclinic trajectory as in Chavanis \cite{C} and Bors, Sta\'nczy \cite{BSM, BSA}
 \end{itemize}

\end{document}